\documentclass[aps,pra,showpacs,twocolumn,superscriptaddress]{revtex4-1}
\usepackage{amsmath}
\usepackage{amssymb}
\usepackage{graphicx}
\usepackage{epstopdf}
\usepackage{times,txfonts}
\usepackage{easybmat}
\usepackage{mathtools}

\usepackage[bookmarks=false]{hyperref}
\hypersetup{colorlinks=true,citecolor=blue,linkcolor=blue,urlcolor=blue,pdfstartview=FitH,bookmarksopen=true}
\usepackage{color,soul}

\begin{document}
\title{Gate Teleportation-based Universal Blind Quantum Computation}
\author{Xiaoqian Zhang}
\email{zhangxq67@mail.sysu.edu.cn}
\affiliation{School of Physics and State Key Laboratory of Optoelectronic Materials and Technologies, Sun Yat-sen University, Guangzhou 510000, China}

\date{\today}

\pacs{03.67.Lx, 03.67.Pp, 03.65.Vf}

\begin{abstract}
  Blind quantum computation (BQC) allows that a client who has limited quantum abilities can delegate quantum computation to a server who has advanced quantum technologies but learns nothing about the client's private information. However, it still remains a challenge to directly encrypt quantum algorithms in circuits model. To solve the problem, we propose GTUBQC, the first gate teleportation-based universal BQC protocol. In this paper, we consider a scenario where a trusted center is responsible for preparing initial states, a client with the ability to perform X, Z does not require any quantum memory, and two servers conducting UBQC (universal BQC) and Bell measurements. GTUBQC can hide the universal quantum gates by encrypting the rotation angles, because arbitrary unitary operation can be decomposed into a combination of arbitrary rotation operators. We prove the blindness and correctness of GTUBQC, and apply our approach to other types of computational tasks, such as quantum Fourier transform.
\end{abstract}

\maketitle

\section{Introduction}
Quantum cloud computing will be very popular with common people (called clients) when a first generation quantum computers come out in the style of `cloud'. More and more people have the demands of quantum computation, however they cannot afford to purchase quantum computers and only have limited quantum technologies. Generally speaking, only some governments and large-scale companies (called servers) have the abilities to purchase and utilize quantum computers. Therefore, a compromise method is that clients can delegate their quantum computation to servers, but how to keep clients' secrets? Fortunately, blind quantum computation (BQC) has been proposed to solve this problem in time \cite{1Broadbent09,Barz12,2Morimae13,3Li14,4Sheng15,5Morimae2012,8Morimae2015,Fitz,Childs2005,22Fisher2014,Broa15}. In BQC, a client with limited quantum technologies delegates her quantum computation to servers, who have full-advanced quantum computers without sacrificing the privacy of her inputs, outputs and quantum algorithms. A. Broadbent \emph{et al.} \cite{1Broadbent09} in 2009 firstly implemented an universal BQC protocol by measuring on blind $m\times n$ dimensional brickwork states, where the client has the abilities to prepare single qubits randomly chosen from a finite set $\{|\pm_\theta\rangle=(|0\rangle+e^{i\theta}|1\rangle)/\sqrt{2}|\theta=0, \frac{\pi}{4}, \frac{2\pi}{4}, \ldots, \frac{7\pi}{4}\}$. Subsequently, S. Barz \emph{et al.} \cite{Barz12} exploited the conceptual framework of measurement-based quantum computation to realize an experimental demonstration ensuring the privacy of quantum inputs, computations, and outputs. After that, double-server and triple-server BQC protocols were proposed in refs. \cite{2Morimae13,3Li14,4Sheng15}. Based on blind topological states \cite{5Morimae2012}, BQC protocol for some single-qubit gates can be realized. The BQC protocol is a concrete fault-tolerant scheme and the error threshold is explicitly calculated. Additionally, an universal BQC can be implemented based on Affleck-Kennedy-LiebTasaki (AKLT) state \cite{8Morimae2015} including blind Z rotation, blind X rotation and controlled-Z followed by blind Z-rotations.

It is obviously that quantum entanglement \cite{Bandyopa11,Shor94,Rausse01,Seth96} plays a key role in measurement-based BQC, moreover, it has many important applications such as quantum nonlocality \cite{Bandyopa11}, quantum error correction \cite{Shor94}, quantum computing \cite{Rausse01} and quantum simulation \cite{Seth96}. Therefore, we have investigated the latest entangled qubits numbers in different experimental physical systems: the largest entangled states are twenty entangled trapped ions \cite{Friis18}, ten entangled photonic qubits \cite{Lin16} and ten entangled superconducting qubits \cite{Song17}. In brief, it is still a challenge to manipulate the number of experimentally controlled single photons such as the brickwork state despite of the rapid development of linear optics technologies \cite{Pan12}.

In 2005, A.M. Childs \cite{Childs2005} first proposed blind quantum computation based on circuits, where the client Alice has the abilities to store quantum states and transmit her qubits, and the server Bob can perform universal quantum computation. K.A.G. Fisher \emph{et al.} \cite{22Fisher2014} realized quantum computation X, Z, H, S, R, CNOT on encrypted quantum states. A. Broadbent \cite{Broa15} introduced an entanglement-based protocol such that it only needs multiple auxiliary qubits or two-way quantum communication. By learning these works, we find that it is still an open problem to hide quantum gates by one-time-pad in BQC based on circuits model. That is to say, if these gates in circuits model can be encrypted similarly to gates in measurement-based BQC model, then the blindness can be achieved perfectly.

In this paper we solve this open problem. We propose the first gate teleportation-based universal blind quantum computation (GTUBQC) protocol, where universal gates set H, T, CNOT are considered. Since arbitrary unitary operators can be decomposed into the combination of rotation operators, gates H, T and CNOT can be concealed by randomly encrypting rotation angles without affecting the quantum computing. In our GTUBQC protocol, there are four participants: a trusted center, a client Alice and two servers Bob1 and Bob2. The trusted center takes responsible for generating resource states and sends qubits to Alice. Alice is an almost classical client because she only needs to perform X, Z operations. Two servers Bob1 and Bob2 are asked to do rotation operations and Bell measurements. GTUBQC ensures that all quantum outputs are at the client's side and the client only needs to detect whether servers honestly return correct measurement outcomes or not. We not only give the proofs of correctness and blindness, but also apply our GTUBQC scheme to realize blind quantum Fourier transform (BQFT) \cite{Marq10,Nam14,Lid17,Moore06}.

The rest of this paper is organized as follows. We present the gate teleportation-based universal blind quantum computation (GTUBQC) protocol in Sec. \ref{sec:jud}. Then we show the analyses and proofs of blindness and correctness, the comparison between measurement-based BQC and gate teleportation-based BQC, and the application in quantum Fourier transform in Sec. \ref{sec:ana}. At last, our conclusions are given in Sec. \ref{sec:con}.

\section{Gate Teleportation-based UBQC protocol}
\label{sec:jud}
\emph{Preliminaries.}---One wants to perform arbitrary sequences of gates from a universal set but unfortunately there will be some by-product Pauli operators in teleportation. For example, sequences of gates $U_3U_2U_1|\psi\rangle$ will be replaced with $P_3U_3P_2U_2P_1U_1|\psi\rangle$ in teleportation where $P_1, P_2, P_3$ are Pauli operators depending on the measurement outcomes. In our GTUBQC protocol, we utilize the rotation operators to implement universal gates, thus we study the Clifford properties of rotation operators \cite{Jozsa05} to get effective quantum computation in teleportation:
\begin{eqnarray}\label{A1}
\begin{array}{l}
\displaystyle R_x(\beta)X=XR_x(\beta), \ \ \ R_x(\beta)Z=ZR_x(-\beta),\\
\displaystyle R_y(\beta)X=XR_y(-\beta),\ \ R_y(\beta)Z=ZR_y(-\beta),\\
\displaystyle R_z(\beta)X=XR_z(-\beta), \ R_z(\beta)Z=ZR_z(\beta).
\end{array}
\end{eqnarray}

\begin{figure*}[!htp]
  \centering
  \includegraphics[width=0.32\textwidth]{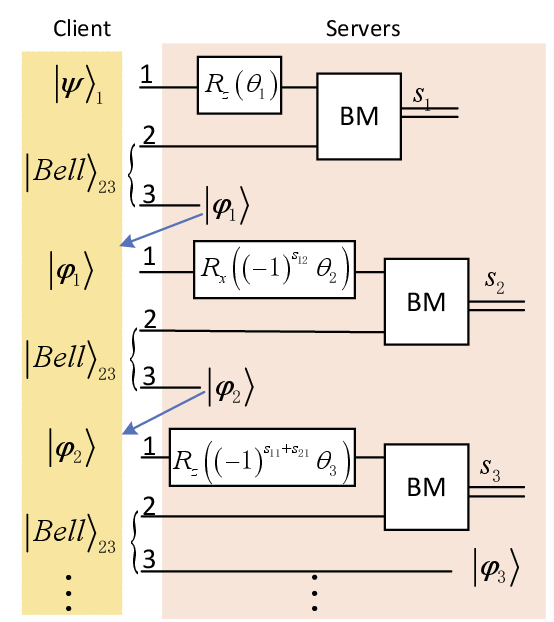}
  \caption{(Colour online) Schematic diagram of single-qubit gates teleportation, where $|\varphi_1\rangle=X^{s_{11}}Z^{s_{12}}R_z(\theta_1)|\psi\rangle_1$, $|\varphi_2\rangle=X^{s_{21}}Z^{s_{22}}R_x((\textnormal{-}1)^{s_{11}}\theta_2)|\varphi_1\rangle$ and $|\varphi_3\rangle=X^{s_{31}}Z^{s_{32}} R_z((\textnormal{-}1)^{s_{22}+s_{12}}\theta_3)|\varphi_2\rangle$. $|Bell\rangle_{23}$ is randomly chosen from $\{|\phi^{\pm}\rangle, |\psi^{\pm}\rangle\}$ and BM denotes Bell measurements. The subscript 1 of $|\psi\rangle_1$ denote the first qubit and the subscripts 2, 3 of $|Bell\rangle_{23}$ denote the second qubit and the third qubit. $s_j\in\{00, 01, 10, 11\}$ represents the measurement outcomes $(j=1, 2, \cdots)$. $s_{j1}, s_{j2}\in \{0,1\}$ are related with by-product operators X and Z respectively. Figure 2 is the same.}\label{F1}
\end{figure*}

In Eq. (\ref{A1}), it is obviously that we need to do adaptive choices of measurements. Besides, the relationship of the rotation angles of $R_x(\cdot), R_y(\cdot), R_z(\cdot)$ are as follows:
\begin{eqnarray}
\begin{array}{l}
\displaystyle R_x(\alpha+\beta)=R_x(\alpha)\cdot R_x(\beta), \\
\displaystyle R_z(\alpha+\beta)=R_z(\alpha)\cdot R_z(\beta), \\
\displaystyle R_y(\alpha+\beta)=R_y(\alpha)\cdot R_y(\beta).
\end{array}
\end{eqnarray}

Another property is that the combination of rotation operators can be used to realize arbitrary unitary {operators \cite{2000MAN}}. For example, there exist $\theta$, $\alpha$, $\beta$ and $\gamma$ such that we obtain z-y-z decomposition as follows:
\begin{small}
\begin{eqnarray}
\setlength{\arraycolsep}{1.2pt}
\begin{array}{l}
\displaystyle U_1=e^{i\theta}R_z(\alpha)R_y(\beta)R_z(\gamma) \\
\displaystyle\quad \ =\left(
  \begin{array}{cc}
  e^{i(\theta-\frac{\alpha}{2}-\frac{\gamma}{2})}cos\frac{\beta}{2} & -e^{i(\theta-\frac{\alpha}{2}+\frac{\gamma}{2})}sin\frac{\beta}{2} \\
  e^{i(\theta+\frac{\alpha}{2}-\frac{\gamma}{2})}sin\frac{\beta}{2} & e^{i(\theta+\frac{\alpha}{2}+\frac{\gamma}{2})}cos\frac{\beta}{2}\\
  \end{array}
\right),
\end{array}
\end{eqnarray}
\end{small}
where $\small
\setlength{\arraycolsep}{1.2pt}
R_y(\beta)=\left(
  \begin{array}{cc}
 cos\frac{\beta}{2} &-sin\frac{\beta}{2} \\
  sin\frac{\beta}{2} &  cos\frac{\beta}{2}\\
  \end{array}
\right),\
R_z(\gamma)=\left(
  \begin{array}{cc}
  e^{\textnormal{-}\frac{i\gamma}{2}} &0 \\
  0 & e^{\frac{i\gamma}{2}}\\
  \end{array}
\right).$

\begin{figure*}[!htp]
  \centering
  \includegraphics[width=0.3\textwidth]{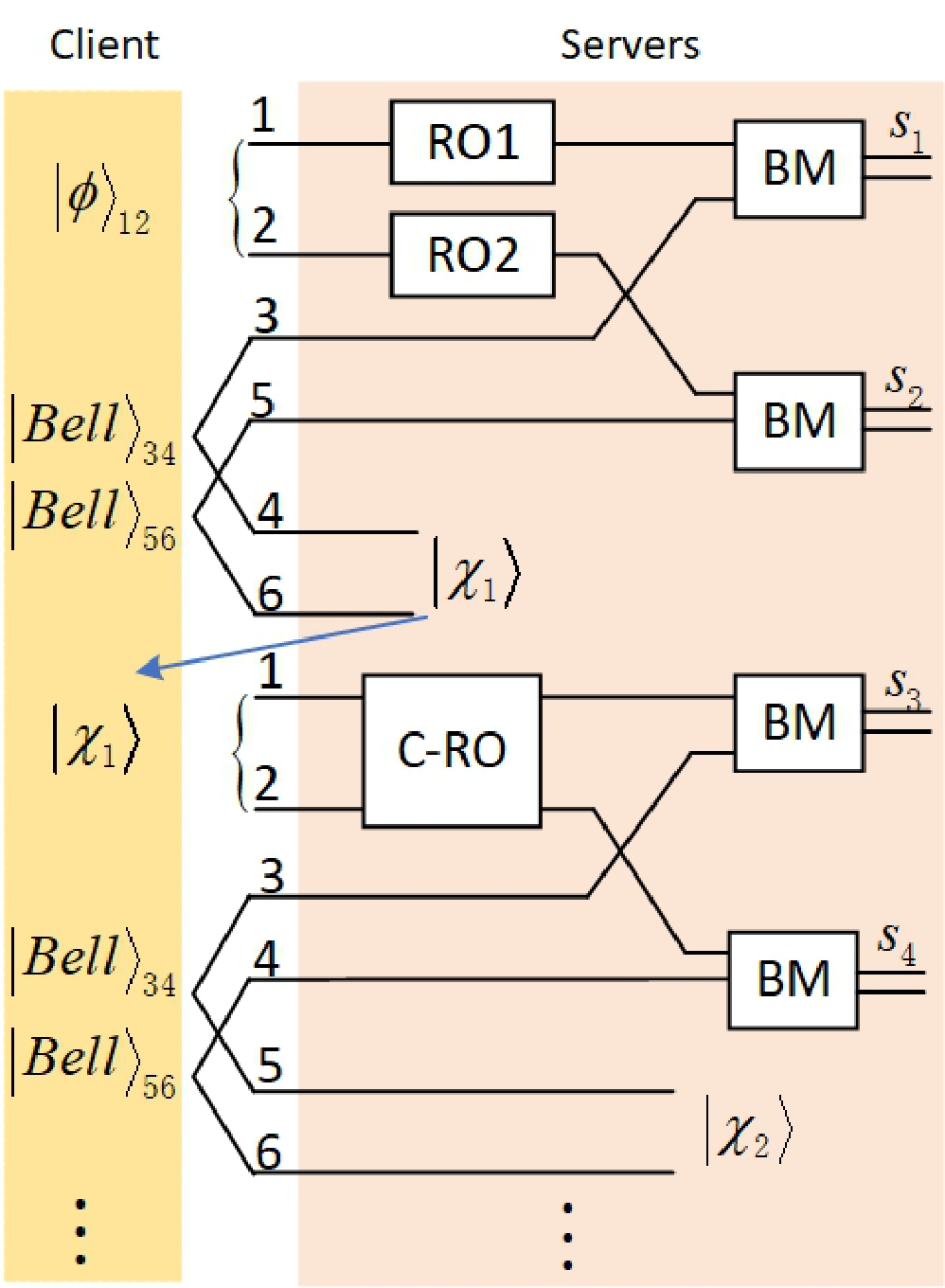}
  \caption{(Colour online) Schematic diagram of double-qubit gates teleportation, where $|\chi_1\rangle=(X^{s_{11}}Z^{s_{12}}\otimes X^{s_{21}}Z^{s_{22}})(RO1\otimes RO2)|\phi\rangle_{12}$ and $|\chi_2\rangle=(X^{s_{31}}Z^{s_{32}}\otimes X^{s_{41}}Z^{s_{42}})C\textnormal{-}RO|\chi_1\rangle$. RO1 and RO2 denote rotation operations. C-RO represents controlled rotation operations.}\label{F2}
\end{figure*}

Here, we give the $z\textnormal{-}y\textnormal{-}z$ decomposition of unitary operators $H, S, Z, T, X, Y$.
\begin{eqnarray}
\begin{array}{l}
\displaystyle H=e^{\frac{i \pi}{2}}R_y(\frac{\pi}{2})R_z(\pi),\ S=e^{\frac{i \pi}{4}}R_z(\frac{\pi}{2}),\ Z=e^{\frac{i \pi}{2}}R_z(\pi), \\
\displaystyle X=e^{\frac{i \pi}{2}}R_y(\pi)R_z(\pi),\ T=e^{\frac{i \pi}{8}} R_z(\frac{\pi}{4}), \ Y=e^{\frac{i\pi}{2}}R_y(\pi),
\end{array}
\end{eqnarray}

\emph{Gate Teleportation.}---The quantum teleportation gates \cite{Jozsa05,2000MAN,Chuang99,Nielsen03,Nielsen04} is similar to quantum teleportation in which two participants previously share halves of a specific two-qubit entangled state, one can teleport a quantum state to the other assisted by classical bits. In Fig. \ref{F1}, we show the process of single-qubit gate teleportation, and we define the relationship between Bell sates and classical bits as $|\phi^+\rangle\leftrightarrow 00, |\psi^+\rangle\leftrightarrow10, |\phi^-\rangle\leftrightarrow01, |\psi^-\rangle\leftrightarrow11$. After every gate teleportation, the by-product operators are $X^{s_{j1}}Z^{s_{j2}}$ and $s_{j1}s_{j2}=s'_j\oplus s_j$, where $s'_j$ is relevant to initial Bell states and $s_j$ is Bob1's (or Bob2's) measurement outcome. We give the detailed teleportation process of rotation operations in Appendix B. The double-qubit gate teleportation is presented in Fig. \ref{F2}.

Next, we show how we can obtain the useful quantum computation by adjust the Pauli operators positions adaptively (See Fig. \ref{F1}) so that these undesirable by-products operators X, Z can be removed easily.
\begin{small}
\begin{eqnarray}
\begin{array}{l}
\displaystyle X^{s_{31}}Z^{s_{32}}R_z((-1)^{s_{11}+s_{21}}\theta_3)X^{s_{21}}Z^{s_{22}}R_x((-1)^{s_{12}}\theta_2) X^{s_{11}}Z^{s_{12}}R_z(\theta_1)|\psi\rangle\\
\displaystyle =X^{s_{31}+s_{21}+s_{11}}Z^{s_{32}+s_{22}+s_{12}}R_z(\theta_3)R_x(\theta_2)R_z(\theta_1)|\psi\rangle
\end{array}
\end{eqnarray}
\end{small}
\noindent where $XZ=-ZX$ and a global phase is ignored.

Subsequently, we present our GTUBQC protocol, where gates H, T and CNOT are decomposed into combinations of (controlled) rotation operators. It is convenient to perform computation since (controlled) rotation operators have the Clifford properties.

\begin{figure*}[!htp]
  \centering
  \includegraphics[width=0.45\textwidth]{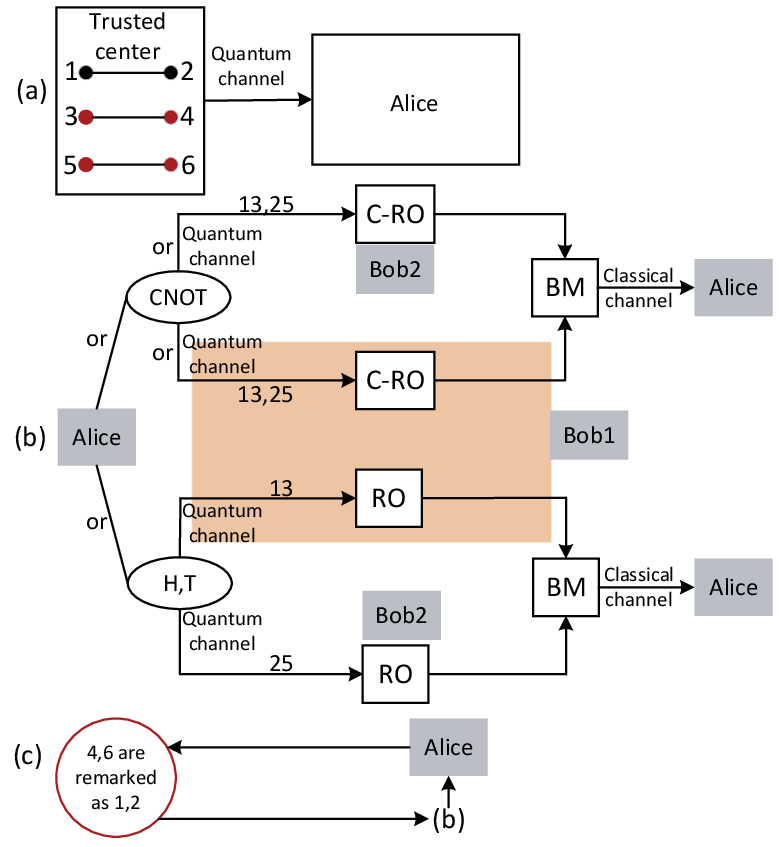}
  \caption{(Color online) Schematic diagram of GTUBQC protocol, where red dots belong to Bell states and black dots belong to arbitrary double-qubit states. RO denotes rotation operations. \textbf{(a)} The trusted center prepares initial states and sends to Alice. \textbf{(b)} Alice performs the computation protocol. \textbf{(c)} Alice and servers repeat the process in \textbf{(b)} until the computation is completed.}\label{F3}
\end{figure*}

\begin{figure*}[!htp]
  \centering
  \includegraphics[width=0.45\textwidth]{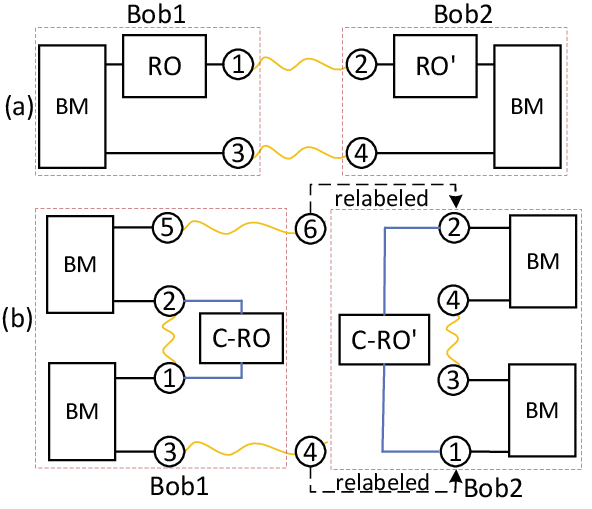}
  \caption{(Colour online) Schematic diagram of servers' honesty test. \textbf{(a)} When the operation is rotation operation (RO), Bob1 and Bob2 share two Bell states. According to Eq. (6), Bob2 performs related operations (RO') and Bell measurements. \textbf{(b)} When the operation is controlled rotation operation (C\textnormal{-}RO), Bob1 and Bob2 share four Bell states. Bob2 performs related controlled operations ((C\textnormal{-}RO')) and Bell measurement.}\label{F4}
\end{figure*}

\emph{GTUBQC protocol.}---In our GTUBQC protocol, there is a trusted center who prepares initial states, a client Alice who is almost classical and two servers Bob1 and Bob2 who perform universal quantum computations and do Bell measurements. One advantage of our GTUBQC is that all qubits are unidirectional transmission, that is, from Alice to Bob1 and Bob2. Servers only need to return classical measurement outcomes. In the following, we give the concrete process of our GTUBQC protocol including \emph{computation process} (See Fig. \ref{F3}) and \emph{test process} (See Fig. \ref{F4}). Alice can implement any one of the two processes at any stage.

\emph{Computation process.}---1) A trusted center prepares enough initial states $|\varphi\rangle$, $|\phi^\pm\rangle$ and $|\psi^\pm\rangle$. As target computational states, qubits 12 belong to $|\varphi\rangle$ which are arbitrary double-qubit states. In teleportation, qubits 34 belonging to Bell states \{$|\phi^\pm\rangle, |\psi^\pm\rangle$\} are as assisted states, the same as qubits 56. Some Bell states are used to detect the correctness of two servers' measurement outcomes in \emph{test protocol}. Then the trusted center sends all initial states to Alice. Alice wants to realize universal quantum computation by the set H, T, CNOT.

2) According to the target computation, Alice can choose rotation operations or controlled rotation operations in the current step. If the computation is a rotation operation, she will send qubits $13$ to Bob1 (or Bob2) and $25$ to Bob2 (or Bob1). Subsequently, Alice sends classical encrypted angles $\theta'_j=r_{j1}\pi+(\textnormal{-}1)^{r_{j2}}\theta_j+\xi_j$ to Bob1 or Bob2, where $\theta_j\in\{\frac{\pi}{4}, \frac{\pi}{2}, \pi\}$, $\xi_j\in\{0, \frac{\pi}{4}, \cdots, \frac{7\pi}{4}\}$, $r_{j1}, r_{j2}\in\{0,1\}$. The value $(\textnormal{-}1)^{r_{j2}}\theta_j$ are actual rotation angles, and $\xi_j$ randomizes the angle $(\textnormal{-}1)^{r_{j2}}\theta_j$. $r_{j1}\pi$ can encrypt the quantum outputs in gate teleportation so that all quantum states are private for servers in the whole protocol, and ${r_{j2}}$ is related with the number of X or Z at the right side of the current operator.

However, if controlled rotation operations need to be performed, Alice will randomly sends qubits $12,35$ to Bob1 (or Bob2). Moreover, Alice randomly sends two qubits to Bob2 (or Bob1) to avoid servers' suspicions. After performing computation, servers do Bell measurement and return classical measurement outcomes to Alice respectively. Here, we show the relationship of encrypted angles and original angles as follows:
\begin{eqnarray*}
R_\phi(\theta'_j)=R_\phi(\xi_j)R_\phi(r_{j1}\pi)R_\phi((\textnormal{-}1)^{r_{j2}}\theta_j),
\end{eqnarray*}
where $\phi=x, y, z$ and $R_x(\pi)= iX$, $R_y(\pi)= XZ$, $R_z(\pi)= iZ.$ Since $R_x(\cdot)$, $R_y(\cdot)$, $R_z(\cdot)$ don't commute with each other except some special cases, these undesired rotation operations affect the primitive blind quantum computation. To eliminate the undesired influence,
we adopt operators $R_\phi(\pi-\xi_j)$ or $R_\phi(2\pi-\xi_j)$. For example, for rotation operator $R_x(\cdot)$, we have
\begin{eqnarray*}
\begin{array}{l}
\displaystyle R_x(\pi-\xi_j)R_x(\theta'_j)=R_x(\pi-\xi_j)R_x(\xi_j)R_x(r_{j1}\pi)R_x((\textnormal{-}1)^{r_{j2}}\theta_j)\\
\displaystyle \qquad\qquad\qquad\ \ \ =iXR_x(r_{j1}\pi)R_x((\textnormal{-}1)^{r_{j2}}\theta_j),\\
\displaystyle \qquad\qquad\qquad\quad\ =-R_x(r_{j1}\pi)R_x((\textnormal{-}1)^{r_{j2}}\theta_j).
\end{array}
\end{eqnarray*}

3) Repeat steps 1-2), until the computing is completed. In the end, Alice performs Pauli operations X and Z to recover the quantum outputs states.

\emph{Test process.}---To test whether servers honestly return Bell measurement outcomes, we utilize entanglement swapping technology to test for rotation operators and controlled rotation operators gate. Note that, all initial states are Bell states.

Firstly, we show that the principles that Bell states remains unchanged after performing some rotation operation such that measurement outcomes of entanglement swapping can be predicted. For any one of Bell states, the rotation operations on the qubit 1 and on the qubit 2 are as follows:
\begin{eqnarray*}
\setlength{\arraycolsep}{1pt}
\begin{array}{l}
\displaystyle |\phi^+\rangle_{12}:R_z(\theta)_1R_z(\textnormal{-}\theta)_2\ or \ R_x(\theta)_1R_x(\textnormal{-}\theta)_2\ or \ R_y(\theta)_1R_y(\theta)_2,\\
\displaystyle |\phi^-\rangle_{12}:{R_z(\theta)_1R_z(\textnormal{-}\theta)_2\ or\ R_x(\theta)_1R_x(\theta)_2\ or\ R_y(\theta)_1R_y(\textnormal{-}\theta)_2},\\
\displaystyle |\psi^+\rangle_{12}:{R_z(\theta)_1R_z(\theta)_2\ or\ R_x(\theta)_1R_x(\textnormal{-}\theta)_2\ or\ R_y(\theta)_1R_y(\textnormal{-}\theta)_2},\\
\displaystyle |\psi^-\rangle_{12}:{R_z(\theta)_1R_z(\theta)_2\ or\ R_x(\theta)_1R_x(\theta)_2\ or\ R_y(\theta)_1R_y(\theta)_2}.
\end{array}
\end{eqnarray*}
The controlled rotation operations are the same.

If Alice performs the test protocol of rotation operations (See Fig. \ref{F4}(a)), Alice sends qubits 13 to Bob1 and 24 to Bob2. Suppose the test Bell states are $|\phi^+\rangle$, Alice asks Bob1 to perform a rotation operation such as $R_z(\theta)$ on qubit 1. Then Bob1 does Bell measurements on qubits 13 and returns the measurement outcomes. To predict the result of entanglement swapping, Alice asks Bob2 to perform $R_z(\textnormal{-}\theta)$ on qubit 2. Similarly, Bob2 does Bell measurements on qubits 24 and returns measurement outcomes. If Bob1 and Bob2 are honest, the outcomes are the same as the expected outcomes and the protocol continues. Otherwise, the protocol is aborted.

However, if Alice performs the test protocol of controlled rotation operations (See Fig. \ref{F4}(b)). Suppose the test Bell states are also $|\phi^+\rangle$, Alice sends 12, 35 to Bob1. Alice asks Bob1 to perform a controlled rotation operation on qubits 12 such as $C\textnormal{-}R_z(\theta)$, and then do Bell measurements on qubits 13, the same as 25. After this, Alice relabels qubits 46 to 12 and sends 12, 34 to Bob2, where Bob2's Bell state 34 are different from the Bell state containing Bob1's qubit 3. Bob2 performs $C\textnormal{-}R_z(\textnormal{-}\theta)$ on qubits 12 and do Bell measurement on 13, the same as 25. Then he returns the measurement outcomes. It is obvious that the entanglement swapping is performed among four Bell states. If these measurement outcomes are related, the test is passed. Otherwise, the protocol is aborted.

The \emph{computation process} and \emph{test process} can be implemented arbitrarily. By \emph{computation process} and \emph{test process}, Alice successfully completes the universal blind quantum computation assisted by Bob1 and Bob2. Note that, Bob1 and Bob2 can communicate with each other, but they still don't know Alice's secret information. Since Alice carries out some tricks in the process of delegating quantum computation. In our proposed protocol, Alice can delegate rotation operators belonging to the same gates to different servers with discontinuous way, but this does not affect the whole computing.

\section{Analyses, comparisons and applications}
\label{sec:ana}
\emph{Analyses}---Firstly, we analyse and prove the correctness and blindness of our GTUBQC protocol.

\emph{Correctness.} If Alice and Bob1, Bob2 follow the steps of our GTUBQC protocol, then all Bell measurement outcomes and quantum outputs are correct.

\emph{Proof:}
1) In Fig. \ref{F1}, these operators are propagated by Pauli operators $X, Z$ combined with Eqs. (1-2), the correctness can refer to Eq. (5). Next, we prove that the encrypted angles don't affect the final quantum computation. The encrypted rotation operators are $\theta_j'=r_{j1}\pi+(\textnormal{-}1)^{r_{j2}}\theta_j+\xi_j$ where $r_{ji}\in\{0,1\}, j=1, 2, 3, i=1, 2$. For example, suppose $\theta'_3=\pi-\theta_3+\xi_3$, $\theta'_2=\theta_2+\xi_2$, $\theta'_1=\pi+\theta_1+\xi_1$, we choose $ R_x(\pi-\xi_3), R_z(\pi-\xi_2), R_x(\pi-\xi_1)$ to remove the influence of undesired rotation operations as follows:
\begin{eqnarray*}
\begin{array}{l}
\displaystyle R_x(\pi-\xi_3)R_x(\theta'_3)R_z(\pi-\xi_2)R_z(\theta'_2)R_x(\pi-\xi_1)R_x(\theta'_1)\\
\displaystyle =R_x(-\theta_3)(iZ)R_z(\theta_2)R_x(\theta_1)=iZR_x(\theta_3)R_z(\theta_2)R_x(\theta_1).
\end{array}
\end{eqnarray*}
Here, we ignore the by-products X and Z in teleportation.$\Box$

\emph{Blindness (quantum inputs).} Suppose the quantum inputs are arbitrary double-qubit states such that servers' density matrix are maximally mixed. That is, Alice has applied the depolarizing channel from the perspective of servers, so servers get nothing about these initial quantum states.

\emph{Proof:}
Suppose $|\psi\rangle=\alpha|00\rangle+\beta|01\rangle+\delta|10\rangle+\eta|11\rangle$, where $|\alpha|^2+|\beta|^2+|\delta|^2+|\eta|^2=1.$ In our GTUBQC protocol, the initial states are arbitrary double-qubit states which are equivalent to do Pauli operation on state $|\psi\rangle$. Moreover, in every teleportation, all quantum inputs for servers are automatically encrypted by $r_{j1}\pi$ since $R_x(r_{j1}\pi)$, $R_y(r_{j1}\pi)$ and $R_z(r_{j1}\pi)$ are I ($r_{j1}=0$) or a combination of X and Z ($r_{j1}=1$). With the help of the equation $\frac{1}{16}\sum_{j,k,l,m=0}^1(Z_1^kX_1^j\otimes Z_2^lX_2^m) |\psi\rangle\langle\psi|(X_1^j Z_1^k\otimes X_2^mZ_2^l)=\frac{I}{4}$,
we can see that the density matrix is independent of quantum inputs.$\Box$

\emph{Blindness (algorithms and outputs).} The blindness of quantum algorithms and quantum outputs can be proved by Bayes' theorem. 1) the conditional probability distribution of Alice's rotation angles is equal to its priori probability distribution, when servers knows partial classical information and measurement outcomes of any positive-operator valued measures (POVMs) at any time. 2) all quantum outputs are one-time padded to servers.

\emph{Proof:} We firstly analyse the effect of Alice's rotation angles information $\Xi_j=\{\xi_j\}_{j=1}^m$ on Alice's {privacy \cite{5Morimae2012,8Morimae2015}}. Suppose $\Theta'_j=\{\theta'_j\}_{j=1}^m$, $\Xi_j=\{\xi_j\}_{j=1}^m$, $\mathbb{R}_{j1}=\{r_{j1}\}_{j=1}^m$ and $\mathbb{R}_{j2}=\{r_{j2}\}_{j=1}^m$, where $\mathbb{R}_{j1}, \mathbb{R}_{j2}\in\{0, 1\}$ are the random variables chosen by Alice and $\{\Xi_j, \Theta'_j \}\in S=\{\frac{k\pi}{4}\mid k=0, 1, 2, \cdots, 7\}$. Let $\mathbb{A}\in\{1,\cdots,m\}$ be a random variable related with some operation.
Bob1's (or Bob2's) knowledge about Alice's secret angles is given by the conditional probability distribution of $\xi_j$ given by $\mathbb{A}=j$ and $\Theta'_j$. Based on Bayes' theorem, we have
\begin{tiny}
\begin{eqnarray*}
\begin{array}{l}
\displaystyle p(\Xi_j=\{\xi_j\}_{j=1}^m \mid \mathbb{A}=j,\Theta'_j=\{\theta'_j\}_{j=1}^m)\\
\displaystyle =\frac{p(\mathbb{A}=j\mid \Xi_j=\{\xi_j\}_{j=1}^m,\Theta'_j=\{\theta'_j\}_{j=1}^m)p(\Xi_j=\{\xi_j\}_{j=1}^m,\Theta'_j=\{\theta'_j\}_{j=1}^m)}{p(\mathbb{A}=j,\Theta'_j=\{\theta'_j\}_{j=1}^m)}\\
\displaystyle =\frac{p(\mathbb{A}=j\mid \Xi_j=\{\xi_j\}_{j=1}^m,\Theta'_j=\{\theta'_j\}_{j=1}^m)p(\Xi_j=\{\xi_j\}_{j=1}^m)p(\Theta'_j=\{\theta'_j\}_{j=1}^m )}{p(\mathbb{A}=j\mid\Theta'_j=\{\theta'_j\}_{j=1}^m)p(\Theta'_j=\{\theta'_j\}_{j=1}^m)}\\
\displaystyle =p(\Xi_j=\{\xi_j\}_{j=1}^m).
\end{array}
\end{eqnarray*}
\end{tiny}
\noindent The reason $p(\mathbb{A}=j\mid \Xi_j=\{\xi_j\}_{j=1}^m,\Theta'_j=\{\theta'_j\}_{j=1}^m)=p(\mathbb{A}=j\mid\Theta'_j=\{\theta'_j\}_{j=1}^m)$ is that each of two servers undertakes a part of tasks such that they can not know which step they are performing in the computation. It means that the conditional probability distribution of Alice's rotation angles is equal to its priori probability distribution. So our GTUBQC protocol satisfies the first condition 1).

Similarly, we show that it is impossible for Bob1 and Bob2 to know the values of $\{r_{j1}\}_{j=1}^m$ only known by Alice. We can get the conditional probability
\begin{tiny}
\begin{eqnarray*}
\begin{array}{l}
\displaystyle p(\mathbb{R}_{j1}=\{r_{j1}\}_{j=1}^m \mid \mathbb{A}=j,\Theta'_j=\{\theta'_j\}_{j=1}^m)\\
\displaystyle=\frac{p(\mathbb{A}=j\mid \mathbb{R}_{j1}=\{r_{j1}\}_{j=1}^m,\Theta'_j=\{\theta'_j\}_{j=1}^m)p(\mathbb{R}_{j1}=\{r_{j1}\}_{j=1}^m,\Theta'_j=\{\theta'_j\}_{j=1}^m)}{p(\mathbb{A}=j,\Theta'_j=\{\theta'_j\}_{j=1}^m)}\\
\displaystyle =\frac{p(\mathbb{A}=j\mid \mathbb{R}_{j1}=\{r_{j1}\}_{j=1}^m,\Theta'_j=\{\theta'_j\}_{j=1}^m)p(\mathbb{R}_{j1}=\{r_{j1}\}_{j=1}^m)p(\Theta'_j=\{\theta'_j\}_{j=1}^m )}{p(\mathbb{A}=j\mid\Theta'_j=\{\theta'_j\}_{j=1}^m)p(\Theta'_j=\{\theta'_j\}_{j=1}^m)}\\
\displaystyle =p(\mathbb{R}_{j1}=\{r_{j1}\}_{j=1}^m).
\end{array}
\end{eqnarray*}
\end{tiny}
The result shows that the value $\{r_{j1}\}_{j=1}^m$ is independent of $\Theta'_j=\{\theta'_j\}_{j=1}^m$, so our GTUBQC protocol satisfies the second condition 2).$\Box$

\emph{Comparison.}---Now, we discuss the measurement-based UBQC and GTUBQC protocols.

1) In measurement-based UBQC model, every gate needs ten-qubit cluster states and the decomposition only is the combination of $R_z(\cdot)$ and $R_x(\cdot)$. However, it still remains a challenge to generate multi-qubits entangled states in experiment. In GTUBQC protocol, we can randomly choose one of six kinds of decompositions and don't need a large-scale entangled states.

2) In both models, the encrypted form are similar, but they have different senses. In measurement-based UBQC model, for $\theta'_j=r_{j1}\pi+(\textnormal{-}1)^{r_{j2}}\theta_j+\xi_j$, $\xi_j$ represents quantum inputs states $|\pm_{\xi_j}\rangle$ unknown by the servers and $\theta_j$ is an actual measurement angle, while $r_{j1}$ have the same meaning in both models: (quantum) outputs are encrypted. However, in GTUBQC protocol, $\xi_j$ is randomly chosen from a finite set $\{0, \frac{\pi}{4}, \cdots, \frac{7\pi}{4}\}$ such that $\theta_j$ can be mapped to a uniform distribution set, moreover $(\textnormal{-}1)^{r_{j2}}\theta_j$ is an actual and adaptive rotation angle. In our protocol, $\xi_j$ will affect the quantum computation but not in measurement-based UBQC model, thus it should be eliminated by some tricks.

\emph{Application.}---Quantum Fourier transform (QFT) can be realized by some ordered single-qubit gates and double-qubit gates. We study the QFT circuits and give the corresponding blind quantum computation protocol with the help of our GTUBQC protocol.

\begin{figure*}[!htp]
  \centering
  \includegraphics[width=0.4\textwidth]{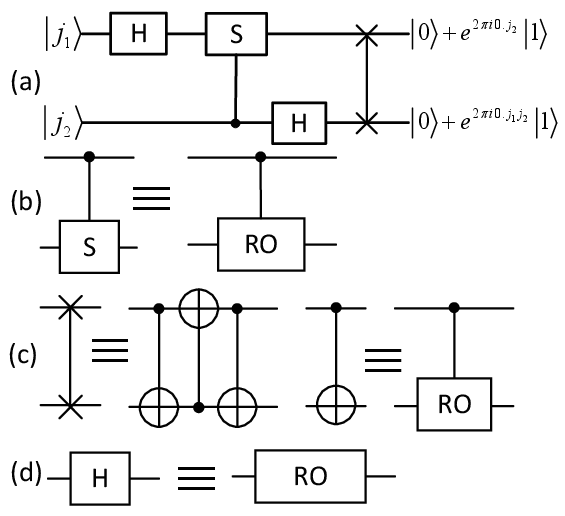}
  \caption{\textbf{(a)} The quantum circuit for two-qubit quantum Fourier transform. \textbf{(b)} The decomposition of gate controlled-S. \textbf{(c)} The decomposition of gate SWAP. \textbf{(d)} The decomposition of gate H.}\label{F5}
\end{figure*}

In Fig. \ref{F5}, we show the original two-qubit QFT circuits in which these gates are decomposed into some basic operations: rotation operations and controlled rotation operations. Gate H can be decomposed into a combination of arbitrary rotation operators. CS and SWAP gates all can be decomposed into a combination of controlled rotation operations. Therefore, the blind two-qubit QFT protocol can be realized by encrypting all rotation angles and randomly splitting computational tasks to two non-communicating servers.

\begin{figure*}[!htp]
  \centering
  \includegraphics[width=0.6\textwidth]{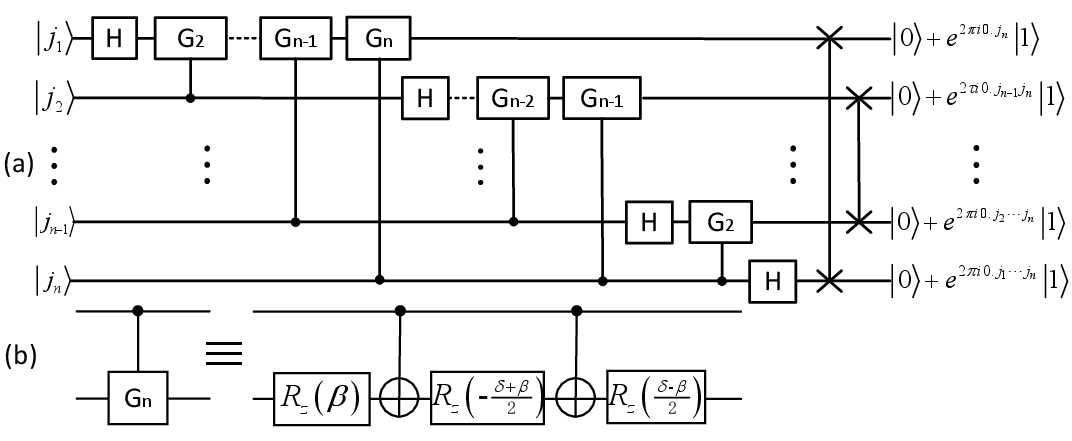}
  \caption{\textbf{(a)} The quantum circuit for multi-qubit quantum Fourier transform. \textbf{(b)} The decomposition of gate $G_n$.}\label{F6}
\end{figure*}

In Fig. \ref{F6}, the multi-qubit QFT circuit is presented and the BQFT also can be realized by the similar method, where gate controlled-$G_n$ can also be decomposed into a combination of rotation operations and CNOT gate. The CNOT gate can be decomposed into a combination of controlled rotation operations. Note that, we use the principle of single-qubit unitary operator: there exist unitary operators A, B, C such that $ABC=I$ and $U=e^{i\alpha}AXBXC$, where $\alpha$ is a global phase factor. Suppose $A=R_z(\beta)R_y(\frac{\gamma}{2}), B=R_y(\textnormal{-}\frac{\gamma}{2})R_z(\frac{\textnormal{-}(\delta+\beta)}{2}), C=R_z(\frac{(\delta-\beta)}{2})$,
$\small
\setlength{\arraycolsep}{1.2pt}
U=\left(
  \begin{array}{cc}
  1   &  0 \\
  0   &  e^{\frac{2\pi i}{2^k}}\\
  \end{array}
\right)$, thus we get $\gamma=0$, $\beta+\delta=\frac{2\pi}{2^k}$ and $\alpha=\frac{\pi}{2^k}$.

\section{Conclusion}
\label{sec:con}
In this paper, we propose an universal blind quantum computation based on gate teleportation which only needs four participants: a trusted center, a client Alice and two servers Bob1, Bob2. The trusted center prepares the initial states and sends to Alice. According to the needs of the computations, Alice respectively distributes qubits to two servers and asks them to perform quantum computation. After that, Bob1 and Bob2 do Bell measurements and return measurement outcomes. In our GTUBQC protocol, there are two parts: computation protocol and test protocol, the former can be used to complete UBQC and the latter can be applied to detect the servers' honesty. The basic universal gates are H, T, CNOT and they can be hidden by encrypting the rotation angles. In conclusion, the quantum computation tasks are allocated to two servers such that they can not reconstruct the quantum computation. Then we not only prove the blindness and correctness but also make a comparison between measurement-based UBQC and GTUBQC protocols. Furthermore, we give the blind protocol of quantum Fourier transform.

\section*{Appendix A: Other combinations of rotation operators}
As we all know, there are six kinds of the combinations of rotation operations. In section 2, we have shown the concrete combinations of rotation operations for some gates in $z\textnormal{-}y\textnormal{-}z$ decomposition. Next, we list two instances of the combinations of rotation operations to realized some gates. For the $y\textnormal{-}x\textnormal{-}y$ decomposition of rotation operators for gates H, S, Z, T, X, Y, we get
\begin{eqnarray*}
\begin{array}{l}
\displaystyle S=e^{\frac{i \pi}{4}}R_y(\frac{-\pi}{2})R_x(\frac{\pi}{2})R_y(\frac{\pi}{2}), \ \ H=e^{\frac{i\pi}{2}}R_x(\pi)R_y(\frac{\pi}{2}),\\
\displaystyle Z=e^{\frac{i \pi}{2}}R_y(\frac{-\pi}{2})R_x(\pi)R_y(\frac{\pi}{2}),\ \ X=e^{\frac{i\pi}{2}}R_x(\pi),\\
\displaystyle T=e^{\frac{i \pi}{8}}R_y(\frac{-\pi}{2})R_x(\frac{\pi}{4})R_y(\frac{\pi}{2}),\ Y=e^{\frac{i \pi}{2}}R_y(\pi).
\end{array}
\end{eqnarray*}
where $\small
\setlength{\arraycolsep}{1.2pt}
R_x(\theta)=\left(
  \begin{array}{cc}
  cos\frac{\theta}{2}   &-isin\frac{\theta}{2} \\
  -isin\frac{\theta}{2} &  cos\frac{\theta}{2}\\
  \end{array}
\right).$

For the $z\textnormal{-}x\textnormal{-}z$ decomposition of rotation operators for gates H, S, Z, T, X, Y, we obtain
\begin{eqnarray*}
\begin{array}{l}
\displaystyle H=e^{\frac{i\pi}{2}}R_z(\frac{\pi}{2})R_x(\frac{\pi}{2})R_z(\frac{\pi}{2}),S=e^{\frac{i \pi}{4}}R_z(\frac{\pi}{2}),\ Z=e^{\frac{i \pi}{2}}R_z(\pi),\\
\displaystyle Y=e^{\frac{i\pi}{2}}R_x(\pi)R_z(\pi),\qquad\quad\ T=e^{\frac{i \pi}{8}}R_z(\frac{\pi}{4}),\  X=e^{\frac{i \pi}{2}}R_x(\pi).
\end{array}
\end{eqnarray*}

\section*{Appendix B: The rotation operations teleportation}
In Fig. 1, suppose $|Bell\rangle_{23}=|\phi^+\rangle_{23}$ and $|\psi\rangle_1=a|0\rangle+b|1\rangle$, where $|a|^2+|b|^2=1$, the process of teleportation is as follows:
\begin{eqnarray*}
\begin{array}{l}
\displaystyle R_z(\theta)|\psi\rangle_1\otimes|Bell\rangle_{23}=e^{\textnormal{-}\frac{i\theta}{2}}(a|0\rangle+e^{i\theta}b|1\rangle)_1\otimes|\phi^+\rangle_{23}\\
\displaystyle \quad=e^{\textnormal{-}i\theta/2}/2[|\phi^+\rangle(a|0\rangle+b e^{i\theta}|1\rangle)+|\phi^-\rangle(a|0\rangle-be^{i\theta}|1\rangle)\\
\displaystyle \qquad\qquad\quad +|\psi^+\rangle(a|1\rangle+be^{i\theta}|0\rangle)+|\psi^-\rangle(a|1\rangle-be^{i\theta}|0\rangle)]_{12,3}
\end{array}
\end{eqnarray*}
If Bob1's measurement outcome is $|\phi^-\rangle$, then Alice obtains the results $s_{j1}s_{j2}=s_j\oplus s'_j=01\oplus00=01$. That is, the by-product operator is $X^0Z^1$. And Alice can obtain $(a|0\rangle+be^{i\theta}|1\rangle)$ from $X^0Z^1(a|0\rangle-be^{i\theta}|1\rangle)$.

\begin{acknowledgments}
This work was supported by the National Natural Science Foundation of China (Grant No. 62005321).
\end{acknowledgments}


\end{document}